\documentclass[preprint]{revtex4-1}
\usepackage{latexsym}

\usepackage{color}

\usepackage{graphicx}
\usepackage{subfigure}
\usepackage{times}
\usepackage{units}
\usepackage{hyperref}
\hypersetup{
  colorlinks = true,
  citecolor = blue,
  urlcolor = blue
}
\newcommand{{\bfr}}{\mbox{\boldmath$r$\unboldmath}}
\newcommand{{\bfv}}{\mbox{\boldmath$v$\unboldmath}}
\newcommand{{\bff}}{\mbox{\boldmath$f$\unboldmath}}
\newcommand{{\bfF}}{\mbox{\boldmath$F$\unboldmath}}
\newcommand{{\bfA}}{\mbox{\boldmath$A$\unboldmath}}
\newcommand{{\bfchi}}{\mbox{\boldmath$\chi$\unboldmath}}

\newcommand{{\cF}}{\mbox{\boldmath${\cal F}$\unboldmath}}
\newcommand{{\cG}}{\mbox{\boldmath${\cal G}$\unboldmath}}
\newcommand{{\cE}}{\mbox{\boldmath${\cal E}$\unboldmath}}
\newcommand{{\cB}}{\mbox{\boldmath${\cal B}$\unboldmath}}
\newcommand{{\cX}}{\mbox{\boldmath${\cal X}$\unboldmath}}
\newcommand{{\cY}}{\mbox{\boldmath${\cal Y}$\unboldmath}}

\begin{document}
\title{A review of Voigt's transformations in the framework of special relativity}
\author{Ricardo Heras}
\email{ricardo.heras.13@ucl.ac.uk}
\affiliation{Department of Physics and Astronomy, University College London, London WC1E 6BT, UK}

\begin{abstract}
\noindent In 1887 Woldemar Voigt published the paper ``On Doppler's Principle,'' in which he demanded covariance to the homogeneous wave equation in inertial reference frames, assumed the invariance of the speed of light in these frames, and obtained a set of spacetime transformations different from the Lorentz transformations. Without explicitly mentioning so, Voigt applied the postulates of special relativity to the wave equation. Here, we review the original derivation of Voigt's transformations and comment on their conceptual and historical importance in the context of special relativity. We discuss the relation between the Voigt and Lorentz transformations and derive the former from the conformal covariance of the wave equation.
\end{abstract}

\maketitle

\section{Introduction}
\noindent
In 1887  Woldemar Voigt published the article \cite{1}: ``On Doppler's Principle,'' which could arguably be considered as the first paper of the relativistic era. In the first part of this article, Voigt demanded covariance to the homogeneous wave equation in inertial reference frames, assumed the invariance of the speed of light in these frames, and obtained a set of spacetime transformations now known as Voigt's transformations. In modern notation these transformations can be written as:
\begin{equation}
 x'=x -vt,\quad t'=t- \frac{vx}{c^2}, \quad y'=  \frac{y}{\gamma},\quad z'=\frac{z}{\gamma},
\end{equation}
where $\gamma=1/\sqrt{1-v^2/c^2}$ is the Lorentz factor. Here we are assuming the standard configuration of special relativity. Voigt's transformations are similar to the well-known Lorentz transformations of special relativity:
\begin{equation}
 x'=\gamma(x -vt),\quad t'=\gamma\bigg(t-\! \frac{vx}{c^2}\bigg), \quad y'=y, \quad z'=z.
\end{equation}
If the right-hand side of Voigt's transformations in equation (1) is multiplied by the Lorentz factor $\gamma$ then the Lorentz transformations in
equation (2) are obtained. Despite the similarity between Voigt and Lorentz transformations, the former are not usually mentioned in standard textbooks \cite{2,3,4,5}. Voigt's transformations have been discussed in old textbooks \cite{6, 7, 8, 9}, old papers \cite{10, 11}, recent articles \cite{12,13,14,15,16,17,18,19,20} and specialized books \cite{21,22,23}.

Some initial comments enlighten the conceptual and historical importance of Voigt's 1887 paper \cite{1}: (i) Voigt derived his transformations by demanding covariance to the homogeneous wave equation under inertial frames, which implied the form invariance of this equation, and this is one application of what would be later known as the first postulate of special relativity. We must say, however, that Voigt didn't explicitly mention the terms ``covariance'' and ``inertial frames.'' He used these concepts in practice; (ii) Covariance of the wave equation carried the invariance of the speed of light, and this would be later known as the second postulate of special relativity. Remarkably,  Voigt inadvertently applied in practice the postulates of special relativity to the wave equation eighteen years before Einstein explicitly and concisely enunciated these postulates \cite{24};(iii) Voigt followed a formal procedure that allowed him to derive a first example of the now known conformal symmetry of spacetime. A general discussion of this symmetry was presented in 1909 by Bateman \cite{25} and Cunningham \cite{26}; (iv) The well-established Newtonian absolute time: $t'=t$ was questioned by Voigt's non-absolute time: $t'=t-vx/c^2$. According to Ives \cite{11} this was the first suggestion that:  ``...a `natural' clock would alter its rate on motion.'' In the same sense, Simonyi \cite{22} has noted that when discarding $t'=t$, Voigt was ``...opening the possibility for the first time in the history of physics to call into question the concept of the absolute time.'' Voigt's non-absolute time was re-introduced in 1895 by Lorentz \cite{27} who called it  ``the local time.''

Interestingly, in a paper devoted to the Doppler effect, Voigt was inadvertently applying the postulates of special relativity nearly two decades before Einstein explicitly and concisely mentioned these postulates. As pointed out by Ernst and Hsu \cite{17}: ``He was very close to suggesting a conceptual framework for special relativity.'' Unfortunately, Voigt's transformations are not usually mentioned in standard textbooks despite the fact that they imply the same transformation law for velocities of special relativity \cite{12}. We believe that two basic reasons are key to understanding why Voigt's transformations have aroused scant interest among authors of standard textbooks: Firstly, these transformations do not form a group \cite{13} which makes them little attractive from a physical point of view.  Secondly, the original derivation of these transformations presented by Voigt \cite{1} is difficult to follow.

In this paper we hope to call attention to Voigt's transformations: (a) by briefly reviewing Voigt's original derivation of these transformations and stressing the scant impact produced by them amongst Voigt's contemporaries; (b) by discussing the relation of these transformations to the Lorentz transformations and (c) by presenting an alternative derivation of Voigt's transformations from the conformal invariance of the d'Alembert operator.

\section {Voigt's 1887 paper}

For a modern reader, Voigt's 1887 paper \cite{1}: ``On Doppler's Principle''  looks like a note, something like a technical report,  rather than a research paper. It does not contain an abstract nor a first paragraph explaining the idea and purpose of the paper. In addition, it does not contain any references.
Many years later, Voigt would explain what he considered to be the basic idea developed in his paper \cite{28}:
\begin{quote} ``...it is about the applications of Doppler's principle, which occur in special parts, though not on the basis of the electromagnetic theory, but on the basis of the elastic theory of light. However, already then some of the same consequences were given, which were later gained from the electromagnetic theory.''
\end{quote}
Voigt mainly focussed his attention on the Doppler effect. The derivation of the transformations in equation~(1) was not his main objective. More specifically, in his 1887 paper, Voigt studied the propagation of oscillating disturbances through an elastic uniform incompressible medium. His basic equation was the homogeneous vector wave equation of the elastic theory of light. As usual in those times, he considered the scalar components of this vector equation. According to Voigt \cite{1}:
\begin{quote}
``It is known that the differential equations for the oscillations of  an elastic incompressible medium read
\end{quote}
\begin{equation}
\frac{\partial^2 u}{\partial t^2}=\omega^2\Bigg(\frac{\partial^2 u}{\partial x^2}+\frac{\partial^2 u}{\partial y^2}
+\frac{\partial^2 u}{\partial z^2}\Bigg),\;\; \frac{\partial^2 {\rm v}}{\partial t^2}=\omega^2\Delta {\rm v},\;\; \frac{\partial^2 w}{\partial t^2}=\omega^2\Delta w,
\end{equation}
\begin{quote}
where $\omega$ is the propagation velocity of the oscillations, or more precisely, the propagation velocity of plane waves with constant amplitude.''
\end{quote}
Voigt wrote $\partial^2 u/\partial t^2=\omega^2\Delta u$ for the first equation in (3), where $\Delta$ is the Laplacian operator. The wave of the elastic incompressible medium is represented by $(u, {\rm v}, w)$. With small changes, we are following the original notation of Voigt's paper.  He also assumed the continuity equation:
$\partial u/\partial x+ \partial {\rm v}/\partial y+ \partial w/\partial z=0.$ Voigt's central idea was to transform equation~(3) from the rest frame to another frame  moving with a constant velocity, and thus finding the formula for the Doppler effect. Without providing any explanation, he wrote a set of linear transformations relating the space coordinates $(x,y,z)$ and the time $t$ measured in the rest frame with the space coordinates $(\xi,\eta,\zeta)$
 and the time $\tau$ measured in the moving frame:
\begin{eqnarray}
\xi&=&xm_1+yn_1+zp_1-\alpha t,\\
\eta&=&xm_2+yn_2+zp_2-\beta t,\\
\zeta&=&xm_3+yn_3+zp_3-\gamma  t,\\
\tau&=&t-(ax+by+cz).
\end{eqnarray}
Voigt introduced 15 unknown constants ($m_1,...,a,..., \alpha,...$) in equations~(4)-(7). Notice that $\gamma$ is not the gamma factor of special relativity. After transforming the first wave equation displayed in (3), in which $u=u(x,y,z,t$), and
using equations~(4)-(7), Voigt obtained a large expression for the transformed wave equation:
 \begin{eqnarray}
\frac{\partial^2 (U)}{\partial \tau^2}\big(1\!-\!\omega^2(a^2\!+\!b^2\!+\!c^2)\big)\!=\!\omega^2\Bigg\{\!\frac{\partial^2 (U)}{\partial\xi^2}\Bigg(\!\!m_1^2\!+\!n_1^2\!+\!p_1^2\!-\!\frac{\alpha^2}{\omega^2}\!\Bigg)\!+\!\frac{\partial^2 (U)}{\partial \eta^2\!}\Bigg(\!m_2^2\!+\!n_2^2\!+\!p_2^2
\!-\!\frac{\beta^2}{\omega^2}\!\Bigg)\!\!+...\nonumber\\
+2\frac{\partial^2(U)}{\partial\eta\partial\zeta}\Bigg(m_2m_3+n_2n_3+p_2p_3-\frac{\beta\gamma}{\omega^2}\Bigg)+ \!...\!\Bigg\}.\quad \quad
\end{eqnarray}
Here $(U)=(U)(\xi,\eta,\zeta,\tau$) is the transformed function corresponding to $u=u(x,y,z,t$). Thus, he compared the transformed wave equation in equation~(8) with the equation
\begin{equation}
\frac{\partial^2 (U)}{\partial \tau^2}=\omega^2\Bigg(\frac{\partial^2 (U)}{\partial \xi^2}+\frac{\partial^2 (U)}{\partial \eta^2}
+\frac{\partial^2 (U)}{\partial \varsigma^2}\Bigg),
\end{equation}
which exhibits the same form as the first equation appearing in (3). Voigt justified the validity of equation~(9) with only four words: ``as it must be'' (or ``da ja sein muss'' in German). Voigt'comparison of equations~(8) and (9) allows him to obtain an algebraic system of 9 equations containing 15 unknown constants ($m_1,...,a,..., \alpha,...$). Some of these equations are: $1-\omega^2(a^2+b^2+c^2) =  m_1^2+n_1^2+p_1^2-\alpha^2/\omega^2,
1-\omega^2(a^2+b^2+c^2)=  m_2^2+n_2^2+p_2^2-\beta^2/\omega^2,...,
m_2m_3+n_2n_3+p_2p_3=\beta\gamma/\omega^2,...$

Voigt's comparison among equations~(8), (9) and the first equation displayed in equation~(3) can be better understood by considering the implied expression
\begin{eqnarray}
\frac{\partial^2 u}{\partial t^2}-\omega^2\Bigg(\frac{\partial^2 u}{\partial x^2}+\frac{\partial^2 u}{\partial y^2}
+\frac{\partial^2 u}{\partial z^2}\Bigg)= {\rm K}\;\Bigg(\frac{\partial^2 (U)}{\partial \tau^2}-\omega^2\Bigg(\frac{\partial^2 (U)}{\partial \xi^2}+\frac{\partial^2 (U)}{\partial \eta^2}+\frac{\partial^2 (U)}{\partial \varsigma^2}\Bigg)\Bigg).
\end{eqnarray}
The parameter ${\rm K}$ in equation~(10) is defined by ${\rm K}\!=\!m_1^2\!+n_1^2\!+\!p_1^2\!-\!\alpha^2/\omega^2$. Equation (10) expresses the covariance of the d'Alembertian of a scalar function defined in both (rest and moving) frames. The typical argument of covariance can be now applied to equation~(10): If the wave equation $\partial^2 (U)/\partial \tau^2\!-\!\omega^2\big(\partial^2 (U)/\partial \xi^2\!+\!\partial^2 (U)/\partial \eta^2\!+\!\partial^2 (U)/\partial \varsigma^2\big)\!=\!0$ is valid in the moving frame then it is valid in the rest frame: $\partial^2 u/\partial t^2\!-\!\omega^2\big(\partial^2 u/\partial x^2\!+\!\partial^2 u/\partial y^2\!+\!\partial^2 u/\partial z^2\big)\!=\!0$ since ${\rm K}\not\!=\!0$. Therefore the covariance of the homogeneous wave equation introduced by Voigt implied the form invariance of this equation under inertial frames, and this is an application of what would be later called by Einstein the first postulate of special relativity. When demanding covariance to the homogeneous wave equation, Voigt assumed (without explicitly specifying it) the constancy of the speed of light $\omega$ in both (rest and moving) frames. This constancy would be later called by Einstein the second postulate of special relativity.

In order to solve his troublesome algebraic system of 9 equations with 15 unknowns, Voigt made an assumption \cite{1}: ``supposed that $\alpha, \beta$ and $\gamma$ [the components of the relative velocity between the rest and moving frames] are given, then we have 12 available constants, so we can arbitrarily use three of them.'' From this point the reading of Voigt's paper becomes somewhat difficult to follow. He re-defined variables and made a number of additional assumptions with the purpose of finding the solution for his troublesome algebraic system.  In essence, Voigt restricted his calculations to work in the standard configuration in which the space coordinates $(x_1,y_1,z_1)$ and the time $t$ are associated with the rest frame and the space coordinates $(\xi_1,\eta_1,\zeta_1)$ and the time $\tau$ are associated with the moving frame. He also wrote the $x-$component of the inter-frame velocity as $\alpha=\chi.$ At the end of his calculations, Voigt obtained the following set of transformations
 \begin{equation}
 \xi_1=x_1 - \chi t,\quad \eta_1=y_1q,\quad \zeta_1=z_1q,\quad \tau=t-  \frac{\chi x_1}{\omega^2},
\end{equation}
where $q\!=\!\sqrt{1-\chi^2/\omega^2}$. If we identify $(\xi_1,\eta_1,\zeta_1)$ with $(x',y',z')$; the time $\tau$ with the time $t'$;  $(x_1,y_1,z_1)$ with $(x,y,z)$ and $q$ with $1/\gamma$ (where $\gamma$ is the Lorentz factor) then we can see that the transformations in equation~(11) are the same as those given in equation~(1). Voigt then proceeded to generalize equation~(11) to be valid for an inter-frame velocity in three dimensions. The remainder of the paper was devoted to a study of the Doppler effect. Because our main concern here are the transformations in equation~(11), or equivalently, those in equation~(1), and not the discussion of the Doppler effect, let us stop here our review of Voigt's 1887 paper.

Let us re-construct the original derivation of Voigt's transformations following a modern and simplified approach. We will see that what Voigt really did in his 1887 paper was to discover one first example of the conformal symmetry of spacetime. Consider the scalar field $F(x,y,z,t)$ in the frame $S$ satisfying the homogeneous wave equation:
\begin{equation}
\frac{\partial^2 F}{\partial x^2}+ \frac{\partial^2F}{\partial y^2}  + \frac{\partial^2F}{\partial z^2} -\frac{1}{c^2}\frac{\partial^2F}{\partial t^2}=0.
\end{equation}
Consider the following set of transformations connecting the space coordinates $(x',y',z')$ and the time $t'$ in the frame $S'$ with the space coordinates $(x,y,z)$ and the time $t$ in the frame $S$:
 \begin{equation}
 x'=k_1x -vt,\quad t'=t-k_2x,\quad y'=k_3y,\quad z'=k_4z,
\end{equation}
where $k_1,k_2, k_3$ and $k_4$ are constants to be determined and $v$ the relative velocity between the frames $S$ and $S'$. From equation~(13) we can derive the transformation laws:
\begin{eqnarray}
\frac{\partial}{\partial y}=k_3\frac{\partial}{\partial y'},\\
\frac{\partial}{\partial z}=k_4\frac{\partial}{\partial z'},\\
\frac{\partial }{\partial x}=k_1\frac{\partial }{\partial x'}-\!k_2\frac{\partial }{\partial t'},\\
\frac{\partial }{\partial t}= \frac{\partial }{\partial t'}-\!v\frac{\partial }{\partial x'}.
\end{eqnarray}
It follows that
\begin{eqnarray}
\frac{\partial^2}{\partial y^2}=k_3^2\frac{\partial^2}{\partial y'^2},\qquad \qquad\\
\frac{\partial^2}{\partial z^2}=k_4^2\frac{\partial^2}{\partial z'^2},\qquad \qquad\\
\frac{\partial^2}{\partial x^2}=k_1^2\frac{\partial^2}{\partial x'^2}-2k_1k_2\frac{\partial^2}{\partial x'\partial t'} +k_2^2\frac{\partial^2}{\partial t'^2},\\
\frac{\partial^2}{\partial t^2}=\frac{\partial^2}{\partial t'^2}-2v\frac{\partial^2}{\partial t'\partial x'} +v^2\frac{\partial^2}{\partial x'^2}.
\end{eqnarray}
Using these transformation laws, equation~(12) is transformed as follows
\begin{eqnarray}
\Bigg(k_1^2-\frac{v^2}{c^2}\Bigg)\frac{\partial^2 F}{\partial x'^2}+ k_3^2\frac{\partial^2F}{\partial y'^2}  + k_4^2\frac{\partial^2F}{\partial z'^2}-(1-k_2^2c^2)\frac{1}{c^2}\frac{\partial^2F}{\partial t'^2}+ \Bigg(\!\frac{2v}{c^2}-2k_1k_2\!\Bigg)\frac{\partial F}{\partial x'\partial t'} =0,
\end{eqnarray}
where now $F(x',y',z',t')$ is defined in the frame $S'.$ Conformal covariance demands
\begin{equation}
\bigg(k_1^2-\frac{v^2}{c^2}\bigg)= k_3^2=k_4^2= (1-k_2^2c^2),\;\; \frac{v}{c^2}-k_1k_2=0.
\end{equation}
By making use of these relations, equation~(22) becomes
\begin{equation}
\bigg(k_1^2-\frac{v^2}{c^2}\bigg)\Bigg(\frac{\partial^2 F}{\partial x'^2}+\frac{\partial^2F}{\partial y'^2}  + \frac{\partial^2F}{\partial z'^2} -\frac{1}{c^2}\frac{\partial^2F}{\partial t'^2}\Bigg) =0.
\end{equation}
From equations~(12) and (24) we obtain the relation
\begin{eqnarray}
\frac{\partial^2 F}{\partial x^2}+ \frac{\partial^2F}{\partial y^2}  + \frac{\partial^2F}{\partial z^2} -\frac{1}{c^2}\frac{\partial^2F}{\partial t^2}=\Bigg(k_1^2-\frac{v^2}{c^2}\Bigg)\Bigg(\frac{\partial^2 F}{\partial x'^2}+\frac{\partial^2F}{\partial y'^2}  + \frac{\partial^2F}{\partial z'^2} -\frac{1}{c^2}\frac{\partial^2F}{\partial t'^2}\Bigg).
\end{eqnarray}
A solution for the system of algebraic equations displayed in equation~(23) is given by
\begin{equation}
k_1=1,\quad k_2=  \frac{v}{c^2}, \quad k_3= \frac{1}{\gamma}, \quad k_4=\frac{1}{\gamma},
\end{equation}
where $\gamma\!=\!1/\sqrt{1\!-\!v^2/c^2}$. Using equations~(13) and (26), we get Voigt's transformations given in equation~(1). Equation~(25) with $k_1\!=\!1$ can compactly be written as
\begin{equation}
\Box^{2}=\frac{1}{\gamma^{2}}\Box'^{2},
\end{equation}
where $\Box^{2}\!\equiv\!\nabla^2\!-\!(1/c^2)\partial^2/\partial t^2$ is the d'Alembert operator in $S$ and $\Box'^{2}\!\equiv\!\nabla'^2-(1/c^2)\partial^2/\partial t'^2$ denotes this operator in $S'$. In section 5 we will present an alternative derivation of Voigt's  transformations based on the conformal covariance expressed in equation~(27).

It is pertinent to mention a common and recurrent misunderstanding, which can be found, for example, in a paper by Rott (See reply in Ref.~\cite{16}):
``It is undisputed that Voigt discovered in 1887 the invariance of the wave equation with respect to the transformation that is named today after Hendrik Lorentz.'' Strictly speaking,  Voigt's transformations (equation~(1)) are not equivalent to the Lorentz transformations (equation~(2)). Doyle has correctly pointed out (See Doyle W T in Ref.~\cite{16}): ``What Voigt actually did was show that the wave equation is covariant under his transformation ---certainly a physical result.''

Authors of books have also incorrectly identified Voigt's transformations with the Lorentz transformations.
Here are some examples.  When referring to the Lorentz transformations, Whittaker \cite{7} claimed: ``It should be added that the transformation
in question had been applied to the equation of vibratory motions many years
before by Voigt,...'' Regarding the origin of the Lorentz transformation, Sears and Brehme \cite{3} wrote: ``In point of fact W. Voigt (1850-1919) first published the transformation in 1887.'' Similarly, French \cite{4} wrote: ``...the Lorentz transformations had, in essence, been discovered in 1887 by W. Voigt, who in that year published a theoretical paper about the Doppler effect (which can be regarded as the problem of observing a wave motion from different inertial frames).'' O'Rahilly \cite{5} has also incorrectly identified Voigt's transformations with the Lorentz transformations. He pointed out: ``The transformation is usually called by the name of Lorentz. But we wish to point out ,..., that the formula was first explicitly given and fully employed by W. Voigt in 1887.'' On the other hand, Brown correctly pointed out \cite{23}:  ``In his 1887 paper, Voigt showed that coordinate
transformations exist-specifically the Lorentz transformations multiplied by $\gamma^{-1}$—which preserve
the form of the wave equations in the elastic theory of light.... It is unclear precisely how Voigt meant the transformations
to be interpreted, or why the multiplicative factor $\gamma^{-1}$ is what it is."

In the next section we will discuss in detail the misunderstanding of considering equivalent the Voigt and Lorentz transformations.

\section {The scant impact of Voigt's 1887 paper}
We wish to highlight three important aspects of the first part of Voigt's 1887 paper, which could explain its scant impact among physicists of that time: (i) The main purpose of Voigt in his 1887 paper was not to propose a new set of spacetime transformations, which should replace the well-established Galilean transformations, but simply to study the transformation of oscillating disturbances through an elastic incompressible medium and deduce the formula for the Doppler effect; (ii) The process by which Voigt derived a set of transformations that maintained covariance of the wave equation was not discussed. As mentioned above, the only four words used by Voigt to justify this covariance were ``as it must be''; and (iii) He did not provide any physical interpretation of his non-absolute time: $t'=t-vx/c^2$ nor did he say anything about the invariance of the speed of light $c$ in inertial frames. Apparently, Voigt  did not realize the great conceptual importance of his transformations!

Regarding Voigt's ideas developed in his 1887 paper,
Hsu has pointed \cite{21}: ``If the physicists of the time had been imaginative enough, they might have recognized the potential of these ideas to open up a whole new view of physics.'' We believe that a considerable number of physicists of that time were imaginative enough and that they did not recognized the potential of Voigt's 1887 paper because Voigt's presentation in the first part of his paper obscured the significance of his transformations.

Despite the great initial acceptance of special relativity in the beginnings of the 1900's, Voigt did not seem to have been interested in pointing out the introduction of his non-absolute time: $t'=t-vx/c^2$. According to the standard account, Lorentz \cite{27} was the first in introducing this time in 1895 but with the name ``local time.'' The replacement of the absolute time by a non-absolute time was a crucial idea in the construction of special relativity. Poincar\'e \cite{29} recognized the great conceptual importance of the local time by claiming that it was Lorentz's ``most ingenious idea'' (in French: ``L'id\'ee la plus ing\'enieuse a  \'et\'e celle du temps local'').

Moreover, evidence points out that Voigt was not substantially interested in promoting his
transformations among his contemporaries. For example, although Voigt had corresponded with Lorentz since 1883, it does not seem that he mentioned his 1887 paper to Lorentz, at least during the period 1887-1907. It was not until 1908 that Voigt sent his 1887 paper to Lorentz. In a response letter, Lorentz wrote \cite{30}:
\begin{quote}
 ``Of course I will not miss the first opportunity to mention, that the concerned transformation
and the introduction of a local time has been your idea.''
\end{quote}
In a letter addressed to Wiechert in the year of 1911, Lorentz \cite {31} mentioned again the priority of Voigt in the discovery of the Lorentz transformations. It is also interesting to note that Lorentz explicitly pointed out in his book \cite{8} that his space-time transformation were first introduced by Voigt. Lorentz wrote:
\begin{quote}
\noindent ``In a paper `\"{O}ber das Dopplersche Princip,' published in 1887 (G\"{o}tt. Nachr., p. 41) and which to my regret has escaped my notice all these years, Voigt has applied to equations of the form (6)
 (\S\,3 of this book)[$\triangle\!\psi\!-\!(1/c^2)\partial^2\psi/\partial t^2\!=\!0$] a transformation equivalent to the formulae (287) and
(288) [$x'\!=\!kl(x\! -\!wt),t'\!=\!kl(t\!-\!wx/c^2),  y'\!=\!ly, z'\!=\!lz $].  The idea of the transformations used above (and in \S\,44) might therefore have been borrowed from Voigt and the proof that it does not alter the form of the equations for the
free ether is contained in his paper.''
\end{quote}
In this paragraph Lorentz pointed out two things: (i) that his transformations were equivalent to those of Voigt and (ii) that Voigt's transformations do not alter the form of the wave equation. The transformations introduced by Lorentz: $x'\!=\!kl(x\! -\!wt),t'\!=\!kl(t\!-\!wx/c^2),  y'\!=\!ly, z'\!=\!lz $ involve both the ``standard'' Lorentz transformations ($w\!=\!v, l\!=\!1$ and $k\!=\!\gamma$) and the Voigt transformations ($w\!=\!v, l\!=\!1/\gamma$ and $k\!=\!\gamma$). The standard Lorentz transformations transform the scalar wave equation $\Box^2 F\!=\!0$ into $\Box'^2 F\!=\!0$ (a perfect invariance) and Voigt's transformations transform $\Box^2 F\!=\!0$ into $\Box'^2 F/\gamma^2\!=\!0$ (a form of covariance). In other words: the homogeneous scalar wave equation is \emph{invariant} under Lorentz transformations and \emph{covariant} under Voigt's transformations. This subtle difference between invariance and covariance of the homogeneous wave equation didn't seem to have been relevant for Lorentz.

Because the Voigt and Lorentz transformations ultimately yield the form invariance of the wave equation (in the case of Voigt's transformations $\Box'^2 F/\gamma^2\!=0$ implies $\Box'^2 F=0$ because $1/\gamma^2\not=0)$, some authors
have incorrectly identified the Lorentz transformations with Voigt's transformations. These authors seem to minimize the difference between covariance and invariance of the wave equation. For example, when discussing the evolution and interpretation of the Lorentz transformations, Pais \cite {32} does not hesitate to say that Voigt was ``the first to write down Lorentz transformations [equations~(2)].'' He points out that wave equations of the type $\Box^2 \phi=0$ ``retains their form if one goes over the new space-time variables [equations (1)].'' He states that ``These [equations~(1)] are the Lorentz transformations [equations~(2)] up to scale factor.'' Pais's statement is incorrect because equations~(1) transform $\Box^2 \phi \!=\!0$ into $\Box'^2 \phi/\gamma^2\!=\!0$ and not into $\Box'^2 \phi\!=\!0$.

It is fair to say that Lorentz always recognized Voigt's 1887 paper. In 1914 he commented on a paper by Poncar\'e and wrote \cite{33}:
 \begin{quote}
``These considerations published by myself in 1904,
have stimulated Poincar\'e to write his article on the
dynamics of electron where he has given my name to
the just mentioned transformation. I have to note as
regards this that a similar transformation has been
already given in an article by Voigt published in 1887
and I have not taken all possible benefit from it.''
\end{quote}
Minkowski didn't seem to have noted some difference between equations~(1) and equations~(2). In a physics meeting of 1908 he claimed (this comment of Minkowski appears the final section (Discussion) of Bucherer's paper given in Ref.~\cite{28}):
 \begin{quote}
\noindent ``I want to add that the transformations, which play the main role in the relativity principle, were first mathematically discussed by Voigt in the year 1887.''
\end{quote}
Voigt's 1887 paper was cited by E. Kohl in 1903 in Annalen der Physik \cite{10}. However, other authors that constructed the special theory of relativity like Einstein and Poincar\'e did not noticed Voigt's 1887 paper.

The fact that Voigt's paper remained unnoticed during the period 1887-1892, was clearly
pointed out by Pauli in his famous 1921 book on relativity \cite{6}. He wrote :
\begin{quote}
``As long ago as 1887, in a paper still written from the point of view of the elastic solid theory of light, Voigt mentioned that it is mathematically convenient to introduce a local time $t'$ into a moving reference frames.... In this way the wave equation $\triangle\phi-(1/c^2)\partial^2\phi/\partial t^2=0$ could be made to remain valid in the moving reference frame, too.  These remarks, however, remained completely unnoticed, and a similar transformation was not again suggested until 1892 and 1895, when H. A. Lorentz published his fundamental papers on the subject.''
\end{quote}
Another 1921 book that mentioned the work of Voigt was that of Kopff \cite{9}.

As a recognition to Voigt, his  1887 paper was reprinted in 1915 in occasion of the tenth anniversary of the principle of relativity (See Ref. \cite {35}).  In connection with this recognition, Doyle writes (See Doyle W T in Ref.~\cite{16}): ``However, Voigt did live to see his [1887] paper chosen
to be reprinted in its entirety in the
Physikalische Zeitschrift (a German
Physics Today of his time) on the
occasion of what the editors called
simply the tenth `birthday celebration of the principle of relativity.'''
In the reprinted version of his 1887 paper (See Ref. \cite{35}), Voigt included some additional comments, the second of them is particularly disconcerting. When referring to his transformations [equation~(11) in the present paper], he wrote:
 \begin{quote}
``This is, except for the factor $q$ which is irrelevant for the application,  exactly the Lorentz transformation of the year 1904.''
 \end{quote}
We know now that the factor $q$ (our $1/\gamma$ in modern notation) is generally significant for applications (Notice Voigt's statement is correct for non-relativistic $(v<<c)$ applications because in this case $q\approx 1$). Apparently, Voigt himself committed the same interpretative mistake as that of Lorentz and Minkowski: he identified his transformations with the Lorentz transformations. Notice that Voigt's comment was made in 1915 when special relativity was already a well-established theory.

\section {The connection between the Voigt and Lorentz transformations}
The relation between the Lorentz and Voigt transformations is transparently established using four-dimensional spacetime notation. This relation is given in Ref. \cite{20} and for completeness we will review it here. Greek indices $\alpha, \beta,  \ldots$ run from 0 to 3; Latin indices $i,j,\ldots$ run from 1 to 3. Coordinates are labeled as $x^{\alpha}\!=\!(x^0, x^1,x^2, x^3)\!=\!(ct,x,y,z)$ in the frame $S$ and $x'^{\alpha}\!=\!(x'^0, x'^1,x'^2, x'^3 )\!=\!(ct',x',y',z')$ in the frame $S'$. Summation convention is adopted.

As is well-known, the Lorentz transformations in equation~(2) can be written as
\begin{equation}
x'^\alpha={\rm \Lambda}_\beta^\alpha \,x^\beta,
\end{equation}
where
\begin{equation}
 {\rm \Lambda}^\alpha_\beta= \left(
\begin{array}{cccc}
\gamma  & -v\gamma/c & 0 & \quad 0 \\
-v\gamma/c & \gamma & 0 &\quad 0 \\
 0 & 0& 1&\quad 0\\
0& 0 & 0 & \quad 1
\end{array}
\right),
\end{equation}
is the Lorentz matrix. Voigt's transformations in equation~(1), on the other hand, can be written as
\begin{equation}
x'^\alpha={\rm V}_\beta^\alpha \,x^\beta,
\end{equation}
where the Voigt matrix is given by
\begin{equation}
{\rm V}^\alpha_\beta=\left(
  \begin{array}{cccc}
    1 & -v/c & 0 & 0\\
    -v/c & 1 & 0 & 0\\
     0 & 0& 1/\gamma & 0\\
    0& 0 & 0 & 1/\gamma\\
  \end{array}
  \right).
\end{equation}
Clearly, the relation between the Lorentz and Voigt matrices is given by
\begin{equation}
{\rm \Lambda}^\alpha_\beta=\gamma{\rm V}_\beta^\alpha,
\end{equation}
i.e., the Lorentz matrix is proportional to the Voigt matrix. Using this proportionality we can infer the properties of the latter from those of the former. The Lorentz matrices satisfy the relation ${\rm \Lambda}^\alpha_\theta{\rm \Lambda}^\theta_\beta=\delta^\alpha_\beta,$ where $\delta^\alpha_\beta$ is the Kronecker delta (See Refs. \cite{36} and \cite{37}).
This means that ${\rm \Lambda}^\theta_\beta$ is the inverse of ${\rm \Lambda}^\alpha_\theta$. This inverse can also be denoted as $({\rm \Lambda}^{-1})^\theta_\beta$. From ${\rm \Lambda}^\alpha_\theta{\rm \Lambda}^\theta_\beta=\delta^\alpha_\beta$ and equation~(32) it follows that
\begin{equation}
{\rm V}^\alpha_\theta \big(\gamma^2{\rm V}^\theta_\beta\big)=\delta^\alpha_\beta.
\end{equation}
Therefore $\gamma^2{\rm V}^\theta_\beta$ can be interpreted as the inverse of ${\rm V}^\alpha_\theta$. This inverse can also be denoted as ${\rm (V^{-1}})^\theta_\beta$. In its explicit form, this inverse transformation reads
\begin{equation}
({\rm V^{-1}})^\theta_\beta=\left(
  \begin{array}{cccc}
    \gamma^2 & -v\gamma^2/c & 0 &\quad 0\;\\
    -v\gamma^2/c & \gamma^2 & 0 &\quad 0\;\\
     0 & 0& \gamma&\quad 0\;\\
    0& 0 & 0 & \quad \gamma\;\\
  \end{array}
  \right).
\end{equation}
The Lorentz matrices are defined to be those satisfying ${\rm \Lambda}^\mu_\alpha\eta_{\mu\nu}{\rm \Lambda}^\nu_\beta=\eta_{\alpha\beta}$, where
\begin{equation}
\eta_{\alpha\beta}=\left(
  \begin{array}{cccc}
    -1 & \;0 &  \;0 & \; 0 \;\\
    \;0  & 1 &  \;0 & 0\\
    \; 0 & 0& 1 & 0\\
    \;0  & 0 & 0 & 1\\
  \end{array}
  \right).
\end{equation}
It follows that the Voigt matrices can be defined as the set of matrices satisfying
\begin{equation}
{\rm V}^\mu_\alpha\gamma^2\eta_{\mu\nu}{\rm V}^\nu_\beta=\eta_{\alpha\beta}.
\end{equation}
We can write equation~(36) in the compact form: ${\rm V^T}\gamma^2\eta{\rm V}=\eta$, where ${\rm V^T}$ is the transpose matrix of ${\rm V}$ \big(notice that ${\rm V^{-1}}=\gamma^2{\rm V}^T\big)$. Despite the close relation between the Lorentz and Voigt matrices, the latter do not form a group \cite{13}.
To show this we consider two Voigt matrices ${\rm V_1}$ and ${\rm V_2}$. We will investigate if their product ${\rm V_1}{\rm V_2}$ is also another Voigt matrix as required by the closure property. We have ${\rm V_1^T}\gamma_1^2\eta{\rm V_1}\!=\!\eta$ and ${\rm V_2^T}\gamma_2^2\eta{\rm V_2}=\eta$. Let ${\rm V_3}\!=\!{\rm V_1}{\rm V_2}.$ Thus ${\rm V_3^T}\gamma_3^2\eta{\rm V_3}\!=\!\big({\rm V_1^T}{\rm V_2^T}\big)\gamma_3^2\eta\big({\rm V_1}{\rm V_2}\big).$ If $\gamma_3\!=\!\gamma_1\gamma_2$ then
\begin{equation}
{\rm V_3^T}\gamma_3^2\eta{\rm V_3}={\rm V_1^T}\gamma_1^2\big({\rm V_2^T}\gamma_2^2\eta{\rm V_2}\big){\rm V_1}={\rm V_1^T}\gamma_1^2\eta{\rm V_1}=\eta.
\end{equation}
From this equation it appears to be that the Voigt matrices satisfy the closure property. But this is not so because the assumption $\gamma_3=\gamma_1\gamma_2$ is incorrect. It can be shown that (See, for example,  Ref. \cite{36} p. 829): $\gamma_3=\gamma_1\gamma_2(1+v_1v_2/c^2).$ The Voigt matrices do not satisfy the closure property and
therefore the Voigt transformations do not form a group since two successive Voigt transformations do not yield another Voigt transformation. This makes them unattractive from a physical point of view because the physical equivalence of the inertial frames is broken.
As pointed out by L\'evy-Leblond \cite{38}: ``The physical equivalence of the inertial frames implies a group structure for the set of all inertial transformations.''

In the low-velocity limit $v \!<\!<\!c$ we have $\gamma\approx 1$ and therefore
\begin{equation}
{\rm \Lambda}^\alpha_\beta\approx{\rm V}_\beta^\alpha.
\end{equation}
The Lorentz and Voigt matrices coincide in the weakly relativistic regime of special relativity. Consequently, the Lorentz and Voigt transformations coincide in this limit. The low-velocity limit of Voigt's transformations take the form
 \begin{eqnarray}
 x'= x -vt,\quad t'=t-\frac{vx}{c^2},\quad y'\approx y,\quad z'\approx z.
\end{eqnarray}
Notice that the time transformation in equation~(37) reduces to the Galilean transformation: $t'=t$ only if the additional condition $ct \!>\!>\!x$ is imposed. In the low-velocity limit we have $\gamma_3\approx\gamma_1\gamma_2$ and then two sets of successive Voigt transformations yield another set of Voigt's transformations, and this satisfies the closure property.  We then conclude that Voigt's transformation approximately form a group in the low-velocity limit $v \!<\!<\!c$. The main objection against Voigt's transformations is seen to disappear when considering low-velocities compared with the speed of light.

\section {Alternative derivation of Voigt's transformations}
In this section we will demand the conformal covariance of the d'Alembert operator expressed in equation~(27) to obtain Voigt's transformations.  This derivation is given in Ref. \cite{20} and for completeness we will review it here again. Consider the standard configuration and equation~(27) expressed as
\begin{eqnarray}
\bigg(\!\frac{\partial}{\partial x}\!-\!\frac{1}{c}\frac{\partial}{\partial t}\!\bigg)\bigg(\!\frac{\partial}{\partial x}\!+\!\frac{1}{c}\frac{\partial}{\partial t}\!\bigg)\!+\!\frac{\partial^2}{\partial y^2}\!+\! \frac{\partial^2}{\partial z^2}\!=\!\frac{1}{\gamma^2}\Bigg(\!\!\bigg(\!\frac{\partial}{\partial x'}\!-\!\frac{1}{c}\frac{\partial}{\partial t'}\!\bigg)\bigg(\!\frac{\partial}{\partial x'}\!+\!\frac{1}{c}\frac{\partial}{\partial t'}\!\bigg)\!+\! \frac{\partial^2}{\partial y'^2}\!+\! \frac{\partial^2}{\partial z'^2}\!\Bigg).
\end{eqnarray}
By assuming linearity for the transformations of derivative operators, we can write
\begin{eqnarray}
\bigg(\frac{\partial}{\partial x}\!-\!\frac{1}{c}\frac{\partial}{\partial t}\bigg) =\frac{A}{\gamma}\bigg(\frac{\partial}{\partial x'}-\frac{1}{c}\frac{\partial}{\partial t'}\bigg),\quad \bigg(\frac{\partial}{\partial x}\!+\!\frac{1}{c}\frac{\partial}{\partial t}\bigg)=\frac{A^{-1}}{\gamma} \bigg(\frac{\partial}{\partial x'}+\frac{1}{c}\frac{\partial}{\partial t'}\bigg),\\
\frac{\partial}{\partial y}=\frac{1}{\gamma}\frac{\partial}{\partial y'},\quad \frac{\partial}{\partial z}=\frac{1}{\gamma}\frac{\partial}{\partial z'}.\qquad\qquad \qquad \qquad \qquad
\end{eqnarray}
Insertion of these quantities into the left-hand side of equation~(40) leads to an identity. The factor $A$ is independent of the derivative operators but can depend on the velocity $v$ and $A^{-1}=1/A$. In order to determine $A$, we demand that the expected linear transformation relating primed and unprimed time-derivative operators should appropriately reduce to the corresponding Galilean transformation \cite{39}: $\partial/\partial t= \partial/\partial t'-v\partial/\partial x'.$ Our demand is consistent with a linear transformation of the general form: $\partial/\partial t= F(v)\big(\partial/\partial t' -v\partial/\partial x'\big),$ where $F(v)$ depends on the velocity $v$ so that $F(v)\!\to\! 1$ when $v<<c$. From this general transformation it follows that if
$\partial/\partial t\!=\!0$ then $\partial/\partial t'\!=\!v\partial/\partial x'$ because $F(v)\neq 0$. Using this result in equations (41) we get
 \begin{eqnarray}
\frac{\partial}{\partial x}=\frac{A}{\gamma}\bigg(\frac{\partial}{\partial x'}\!-\!\frac{v}{c}\frac{\partial}{\partial x'}\bigg),\quad
\frac{\partial}{\partial x}=\frac{A^{-1}}{\gamma}\bigg(\frac{\partial}{\partial x'}\!+\!\frac{v}{c}\frac{\partial}{\partial x'}\bigg).
\end{eqnarray}
By combining these equations we can derive expressions for $A$ and $A^{-1}$,
\begin{equation}
A=\frac {\sqrt{1+v/c}}{\sqrt{1-v/c}},\quad A^{-1}=\frac {\sqrt{1-v/c}}{\sqrt{1+v/c}},
\end{equation}
which can conveniently be written as
\begin{equation}
A=\gamma\bigg(1+\frac{v}{c}\bigg),\quad A^{-1}=\gamma\bigg(1-\frac{v}{c}\bigg).
\end{equation}
Using these relations in equation~(41) we obtain
\begin{eqnarray}
\bigg(\frac{\partial}{\partial x}-\frac{1}{c}\frac{\partial}{\partial t}\bigg)=\bigg(1+\frac{v}{c}\bigg) \bigg(\frac{\partial}{\partial x'}-\frac{1}{c}\frac{\partial}{\partial t'}\bigg),\,\, \bigg(\frac{\partial}{\partial x}+\frac{1}{c}\frac{\partial}{\partial t}\bigg)=\bigg(1-\frac{v}{c}\bigg) \bigg(\frac{\partial}{\partial x'}+\frac{1}{c}\frac{\partial}{\partial t'}\bigg).
\end{eqnarray}
By adding and subtracting equations (46) we can obtain the corresponding transformation laws connecting unprimed and primed operators, which are added to the transformation laws for $\partial/\partial y$ and $\partial/\partial z$, obtaining
\begin{equation}
\frac{\partial}{\partial x}\!=\!\frac{\partial}{\partial x'}-\!\frac{v}{c^2}\frac{\partial}{\partial t'},\quad
\frac{\partial}{\partial t}\!=\! \frac{\partial}{\partial t'}-\!v\frac{\partial}{\partial x'},\quad \frac{\partial}{\partial y}\!=\!\frac{1}{\gamma}\frac{\partial}{\partial y'},\quad
\frac{\partial}{\partial z}\!=\!\frac{1}{\gamma}\frac{\partial}{\partial z'}.
\end{equation}
These relations are the Voigt transformations for derivative operators of the standard configuration. The first two relations in equation~(47) imply coordinate transformations of the form: $x'=x'(x,t)$ and $t'=t'(x,t).$ To find the explicit form of these transformations we can use the first two relations displayed in equation (47) to obtain
\begin{equation}
\frac{\partial x'}{\partial x}= 1,\quad
\frac{\partial x'}{\partial t}= -v,\; \frac{\partial t'}{\partial t}= 1,\quad
\frac{\partial t'}{\partial x}=-\frac{v}{c^2}.
\end{equation}
From the first relation in equation (48) it follows the equation (A): $x'\!=\! x \!+ \!g_1(t),$ where $g_1(t)$ can be determined (up to a constant) by deriving (A) with respect to the time $t$ and using the second relation in (51): $\partial x'/\partial t\!=\!\partial g_1(t)/\partial t\!=\!-v.$ This last equality implies (B): $g_1(t)\!=\!-vt\!+\! x_0,$ where $x_0$ is a constant. From (A) and (B) we obtain
\begin{equation}
x'=x -vt + x_0.
\end{equation}
The third relation in equation~(48) implies (C): $t'\!=\!t\! +\! g_2(x),$ where $g_2(x)$ can be obtained (up to a constant) from deriving (C) with respect to $x$ and using the last relation in equation~(41):
$\partial t'/\partial x\!=\!\partial g_2(x)/\partial x\!=\!-v/c^2$. This last equality implies (D):
$g_2(x)\!= -(v x/c^2)\!+\! t_0,$
where $t_0$ is a constant. From (C) and (D) we conclude
\begin{equation}
t'=t - \frac{vx}{c^2} + t_0.
\end{equation}
 The origins of the frames $S$ and $S'$ coincide at  $t\!=\!t'\!=\!0$. It follows that $x_0\!=\!0$ and $t_0\!=\!0$. In this way we obtain the Voigt transformations for the $x$ and $t$ coordinates of the standard configuration:
\begin{equation}
 x'=x -vt,\quad t'\!=t-   \frac{vx}{c^2} .
\end{equation}
The transformations for the $y$ and $z$ coordinates are easily derived. From the last two relations in equation~(50) we get $\partial y'/\partial y\!=\!1/\gamma$ and $\partial z'/\partial z\!=\!1/\gamma$. They imply $y'\!=(y/\gamma) + \!y_0$ and $z'\!=(z/\gamma) + z_0$, where $y_0$ and $z_0$ are constants, which vanish because the origins of the frames $S$ and $S'$ coincide at the time  $t\!=t'=0$. Thus
\begin{equation}
 y'=\frac{y}{\gamma}, \quad z'= \frac{z}{\gamma}.
\end{equation}
Equations~(51) and (52) are the Voigt transformations of the standard configuration. A direct manipulation of equations~(51) and (52) yields the corresponding inverse transformations \cite{13}:
 \begin{equation}
 x=\gamma^2\big(x'+vt'\big),\quad t=\gamma^2\bigg(t'+  \frac{vx'}{c^2}\bigg), \quad y=\gamma y',\quad z=\gamma z'.
\end{equation}

\section{Conclusion}
In the creation of special relativity, we traditionally find the names of Einstein, Poincar\'e, Lorentz and Larmor. They appear to be the main actors. Voigt is relegated to be a minor player, in the best of cases. But this tradition is not faithful to the history of physics. Although Voigt did not derive the Lorentz transformations, he seems to have been the first in applying the postulates of special relativity to a physical law: he demanded covariance to the homogeneous wave equation with respect to inertial frames and assumed the invariance of the speed of light in these frames, obtaining a set of transformations which introduced the non-absolute time $t'=t-vx/c^2.$

To conclude, we would like to point out that we are in agreement
with Rott \cite{15}, who on the centennial of Voigt's 1887 paper, claimed that Voigt was a ``Relativity's forgotten figure.'' He wrote \cite{15}: ``...this year is also the centennial of a theoretical paper [Voigt's 1887 paper] that is largely forgotten, but has a certain role in the history of the theory of relativity even though it has no documented impact on the actual historical development.''
\section*{Acknowledgements}
I thank James Read for his helpful and insightful comments on this paper.

\end{document}